\documentclass[twocolumn,prd]{revtex4} 
\usepackage[dvipdfmx]{graphicx}

\begin{document}

\title{ Quark-gluon cluster -- a composite entity just due to color confinement}
\author{			Keiichi Akama}
\affiliation{	Department of Physics, Saitama Medical University,
 			 Saitama, 350-0495, Japan}
\date{\today}

\begin{abstract}	
We incorporate confining effects 
	into statistical formulation of quantum chromodynamics (QCD). 
It implies that quark-gluon clusters existed in the early universe, 
	triggering the information accumulation in nature. 
The confinement-driven clustering would explain the reason  
	why the quark-gluon plasma in the heavy-ion collision experiments looks like liquid 
	unlike the QCD expectation.
\end{abstract}

\maketitle

\section{Introduction}

In this paper, we incorporate confining effects 
	into statistical formulation of quantum chromodynamics (QCD). 
This implies emergence of {\it quark-gluon clusters}, 
	i.e.\ composite entities of many quarks and gluons 
	bound just due to color confinement.
It brings us new insights into the quark-gluon plasma (QGP) \cite{QGP}, \cite{QGPrev}  
	in the early and in the present universe, 
	as well as into information accumulation in nature  \cite{information}--\cite{Darwin}.
Taking into account the baryon number conservation in the expanding universe \cite{Weinberg},
	we will claim that such clusters existed in a certain period in the early universe,  
	yielding the first macroscopic inhomogeneity 
	far prior to hadrons.
It will turn out that 
	the clusters held large amounts of information for their own existence,
	initiating the information accumulation long inherited in nature. 
It is known that the ``QGP" observed in the heavy-ion collision experiments today 
	looks strongly interacting like liquid, differing from the naive expectation of QCD \cite{QGPrev}, \cite{sQGP}. 
We will argue that this is because the quarks and gluons 
	are grouped into strongly interacting clusters due to the confining force, 
	and it would microscopically solve the discord above.

The plan of this paper is as follows.
In Chap.\ II, we advocate the quark-gluon cluster as the initiator of information accumulation in nature.
In Chap.\ III, we incorporate confining effects 	into statistical formulation of QCD,
	and derive the quark-gluon cluster and its properties.
In Chap.\ IV, we derive the important consequences stated above,
	and Chap.\ V is devoted to the conclusions.

\section{Quark-gluon cluster}

Here, we remark an important role of the quark-gluon clusters 
	in information accumulation in nature \cite{information}.
We observe that
	composite entities such as molecules, cells, organisms, societies, and species 
	accumulate tremendous information for their own existence \cite{Akama}.
Similarly, stones, mountains, rivers, typhoons, stars etc.\ 
	keep their existence against disturbances
	on the basis of their own and environmental information for their existence.   
Only the composites equipped with appropriate information for existence 
	can appear as the ingredients of existing nature.
The information is refined and accumulated through competition and selection 
	(in the sense of \cite{Darwin}) of general composites for their existence.
Extensive studies in various sciences such as evolution, informatics, history, archaeology, 
	geology, meteorology and cosmology have demonstrated 
	that information \cite{information} in nature is systematically organized and accumulated 
	based on its preceding pieces successively.

Then, what is the origin of the information in nature?
The most fundamental composites around us today are nucleons 
	\cite{composite}--\cite{subquark}.  
A nucleon keeps its existence owing to the facts that 
	i) its constituents are quarks and are subject to binding due to QCD, 
	ii) it has appropriate energy, 
	iii) it is protected by the baryon number conservation,
	and iv) it is in an appropriate environment.
These facts i)--iv) can be taken as the pieces of information \cite{information} 
	for the nucleon to exist and to keep existence.
Based on them, atoms and molecules are formed with additional information,
	and subsequently formed are crystals, molecular systems,
	and further multiple composites like organisms and other structures, 
	with so sophisticated networks of information to guarantee their existence.

Suppose we know the probability distribution of the states of a composite 
	and that without the information to keep its compositeness. 
Then, we can estimate the amount of information a la Shannon 
	by subtracting the entropy from that without it \cite{Shannon}.
For example, 
	for a proton with mass $m_{\rm p}$  
	in a box with the edge $L$ and the temperature $T$, 
	the amount of information for its existence is estimated to be (in the natural unit)
\begin{eqnarray}&&
	I_{\rm p}=\ln[{(em_{\rm p}T)^3 L^6}/{4\sqrt{3}\pi^3}]. \label{Ip}
\end{eqnarray}
In deriving (\ref{Ip}), we assumed
	ideal gas for the proton, and that for three free quarks 
	with mass $m_{\rm q}\sim m_{\rm p}/3$  
	as the system without the information.
In similar ways, we can estimate information amounts of 
	atomic nuclei, atoms, ions, molecules, and so forth,
	and can quantitatively trace the information accumulation in nature \cite{Akama}.

Thus, the origin of information evolution in nature is traced back to the nucleons. 
In this paper, however, we advocate their possible progenitors, the quark-gluon clusters.
We presume that the early universe is filled with
	uniform plasma of fundamental fields \cite{composite} 
	with extremely high temperature and density \cite{braneworld}. 
Within this era, no information was realized,  
	since information makes sense 
	only when its goal and its user exist~\cite{information}.
As the universe cooled, however, quark-gluon clusters
	were formed just due to {color confinement}. 
The confining nature of the quarks and gluons  
	is taken as a piece of information for each cluster to exist. 
The goal of the information is existence of the cluster and the user is the cluster itself.
In fact, its amount is estimated to be huge, as we will see later. 
In further cooling, the clusters reduced into hadrons, 
	and, among them, nucleons survived owing to the information they held.   
Thus, the quark-gluon clusters triggered the information accumulation 
	long inherited in nature up until the present time.

\section{formulation}

To formulate this precisely, let us consider a system of quarks and gluons  
	in a cubic box with the edge length $L$ in a heat reservoir with the temperature $T$.
Then, the probability $p_i$ of the state $i$ with energy $E_i$ 
	and the total number $N_i$ of quarks and gluons is given by
\begin{eqnarray}&&
	p_i=e^{(\mu N_i-E_i)/T}/\Xi \ \ \mbox{with } \Xi={\sum}_i e^{(\mu N_i-E_i)/T},\ \ \ 
\end{eqnarray}
where 
	$\mu$ is the chemical potential.
Here, we only consider the case with $\mu\approx 0$.
We are interested in the average particle number $\langle N\rangle=T{\partial \ln \Xi}/{\partial \mu}$, 
	the average energy $\langle E\rangle=T^2{\partial \ln \Xi}/{\partial T}$, 
	and the Shannon entropy 
\begin{eqnarray}&&
	S\equiv-{\sum}_i p_i \ln p_i={\partial (T\ln \Xi)}/{\partial T}\ \ \label{EPSdef}
\end{eqnarray}
of the system.
Note that the system itself is not necessarily statistical, 
	but can be with finite degrees of freedom like a nucleon or a cluster. 
We do not require thermodynamic limit for the system itself.
Therefore, the averaged quantities can have non-vanishing fluctuations in general. 
If we neglect interactions among particles, the quantum theory 
	implies that $\Xi$ is given by 
\begin{eqnarray}&&
	\Xi=\exp\sum_{\pm,\ \epsilon}{\cal F}_\pm, \ \ 
	{\cal F}_\pm=\pm g_{\pm}\ln(1\mp e^{(\epsilon-\mu)/T}),\ \ \ \label{Z}
\end{eqnarray}
where $\epsilon$ is the energy eigenvalue of the one particle states,
	and $g_+=16$ and $g_-=36$, respectively, are the degrees of freedom 
	of gluons and light quarks (three light flavors are assumed) 
	with energy $\epsilon$.

Now, we incorporate the effects of color confinement of QCD.
According to it, any physical state should form a color singlet.
It is realized by connecting quarks and gluons with appropriate gluon-flux strings,
	which carry extra energy proportional to their lengths. 
In fact, models with linear potential $U(r)=\sigma r$ 
	(with the distance $r$ of particles and a constant $\sigma$) 
	are phenomenologically successful 
	in explaining quarkonium spectra \cite{linearpot}.
Taking mean field approximation, we assume that each particle moves in the potential 
	$U(r)=\sigma r$ in $r\le L$ while $\infty$ in  $r>L$.
We approximate the spectra of the regions $\epsilon> \sigma L$ and $\epsilon\le \sigma L$ 
	by the asymptotic form in each, and neglect the further interactions.

In $\epsilon\gg \sigma L$, 
	the $r$-dependence of the potential is negligible, 
	and hence 
	$\epsilon\approx \pi n/L$ with an integer $n$.
As for the asymptotic form in $\epsilon\le \sigma L$, 
	we need to extrapolate the Schr\"odinger equation with the potential $U(r)=\sigma r$ 
	to the Klein-Gordon equation.
A natural and asymptotically consistent extension is
\begin{eqnarray}&&
	-\nabla^2\psi+(m+\sigma r)^2\psi=\epsilon^2\psi,
\label {KG}
\end{eqnarray}
	where $\psi$ is the one-particle wave function, 
	$m$ is its mass, $\epsilon$ is its energy, 
	and $r$ is the distance from the center of mass in its rest frame. 
This reduces to the Schr\"odinger equation 
	when $m\gg r\sigma$ and $m\gg \epsilon-m>0$.
We assume that the chiral symmetry is restored at some temperature, 
	so that $m\approx 0$ beyond, and many quark-antiquark pairs are produced.
The asymptotic form of the radial part of (\ref{KG}) in $r\rightarrow\infty$
	has the asymptotic spectrum $\epsilon\approx\sqrt{2n\sigma}$ with an integer $n$. 
Since the gluons are expected to be confined in a similar way,
	we use (\ref{KG}) with $m=0$ for gluons with the same $\sigma$ for simplicity.

Using these asymptotic spectra and spherical symmetry, 
	we replace $\sum_\epsilon$ in (\ref{Z}) by $\int 4\pi n^2dn$ and obtain
\begin{eqnarray}&&
	\sum_{\epsilon}{\cal F}_\pm=  
	\int_0^{\sigma L}{\cal F}_\pm \frac{\pi f}{8\sigma^3}\epsilon^{5}d\epsilon
	+\int_{\sigma L}^\infty{\cal F}_\pm	\frac{L^{3}f}{2\pi^2}\epsilon^{2}d\epsilon,
\ \ \ 
\label {int}
\end{eqnarray}
where $f$ is the function of $\epsilon$ to give the deviations from the asymptotic form 
	at the lower energy of each, though we need not to specify it here.
The integrand of (\ref{int}) for $\epsilon\gg\sigma L$ is
	that for the ideal QGP. 
Meanwhile, that for $\epsilon\le\sigma L$ is 
	that in which the $L$ is replaced by $\epsilon/\sigma$
	in that for the ideal QGP (apart from an extra factor $\pi^3/4$). 
Physically, this means that  
	the wave function is localized within $r=\epsilon/\sigma$ 
	owing to the confining potential $U(r)=\sigma r$,
	though the levels are denser by the extra factor.
It is reasonable because the less steep potential wall
	admits the denser levels.
It is a consequence of the confining force.
Hence, the whole color singlet system is also dominated 
	by localized ones which leave open space filled only with photons and leptons. 
In fact, its typical size is estimated below (see (\ref{R})). 
Statistical ensembles of such localized states 
	with certain sets of macroscopic indices represent 
	a localized macroscopic entity.
It is nothing but the ``quark-gluon cluster" 
	defined at the top of this paper.
The cluster is expected to be in equilibrium 
	with plasma of photons and leptons with temperature $T$.
We call this the ``single cluster model''.
The more realistic cases of many interacting clusters will be discussed later 
	on the basis of this simple case.

Let us consider the behaviors of $\langle N\rangle$, 
	$\langle E\rangle$, and $S$ of the system according to (\ref{int}). 
At very high temperature $T\gg \sigma L$, it exhibits the ideal QGP with
\begin{eqnarray}&&
	\langle N\rangle=c_3T^3L^3,  \label{n0}
\\&&
	\langle E\rangle= c_4T^4L^3, \label{EhiT}
\\&&
	S= (4/3)c_4T^3L^3, \label{high}
\end{eqnarray}
where $c_\alpha\equiv
	\Gamma(\alpha)\zeta(\alpha)(g_++(1-2^{1-\alpha})g_-)/2\pi^2$
	with the Riemann zeta function $\zeta(z)$, and hence
	$c_3\approx5.24$ and $c_4\approx15.6$.
At the intermediate temperature $T_{\rm had}(\equiv$ a few TK $)\ll T\ll \sigma L$, 
	where the quark-gluon cluster is formed,
	we have
\begin{eqnarray}&&
	\langle N\rangle
	=(\pi^3/4)c_{6}T^{6}/\sigma^3\approx 0.25\,T_{\rm^{TK}}{}^6,\label{NmidT}
\\&&
	\langle E\rangle
	=(\pi^3/4)c_{7}T^{7}/\sigma^3\approx (0.13{\rm GeV})T_{\rm^{TK}}{}^7,\label{EmidT}
\\&&
	S=(7\pi^3/24)c_{7}T^{6}/\sigma^3\approx 1.7\,T_{\rm^{TK}}{}^6,
	\label{intermediate}
\end{eqnarray}
where $c_6\approx318$, $c_7\approx1892$, we used the empirical value $\sigma=0.16\rm GeV^2$, and
	$T_{\rm^{TK}}=T/{\rm{TK}}$ is the numerical value of $T$ in the unit of TK.
These are independent of $L$ as they should be.
The densities $n_{\rm in}$, $e_{\rm in}$ and $s_{\rm in}$ of
	$\langle N\rangle$, $\langle E\rangle$, and $S$, respectively,
	per unit volume \underline{inside} the cluster 
	are obtained by dividing them by the extension volume
 	$(\epsilon/\sigma)^3$ of each particle
	\underline{before} the integration with $\epsilon$:
\begin{eqnarray}&&
	n_{\rm in}=(\pi^3/4)c_3 T^{3}\approx(3.4{\rm/fm^{3}})T_{\rm^{TK}}{}^3,\label{nin}
\\&&
	e_{\rm in}=(\pi^3/4)c_4 T^{4}\approx(0.87{\rm GeV/fm^{3}})T_{\rm^{TK}}{}^4,
	\label{einmidT}
\\&&
	s_{\rm in}=(\pi^3/3)c_{4} T^{3}\approx(13{\rm/fm^{3}})T_{\rm^{TK}}{}^3.\label{sin}
\end{eqnarray}
They	behave parallel to those of the ideal QGP 
	(obtained by dividing (\ref{n0})--(\ref{high}) by $L^3$),
	though the magnitudes are larger by $\pi^3/4$,
	as is a consequence of the confining effects in QCD.
Comparing (\ref{NmidT})--(\ref{intermediate}) and (\ref{nin})--(\ref{sin}),
	we find that  
	the typical size $R$, 
	the ``energy size'' $R_{\rm E}$ and 
	the ``entropy size" $R_{\rm S}$ of the cluster are given by
\begin{eqnarray}&&
	R\equiv\sqrt[3]{\langle N\rangle/n_{\rm in}}=\sqrt[3]{c_6/c_3}T/\sigma\approx (0.42{\rm fm})T_{\rm^{TK}}, \label{R}
\\&&
	R_{\rm E}\equiv\sqrt[3]{\langle E\rangle/e_{\rm in}}=\sqrt[3]{c_7/c_4}T/\sigma\approx (0.53{\rm fm})T_{\rm^{TK}},\ \ \  \label{RE}
\\&&
	R_{\rm S}\equiv\sqrt[3]{S/s_{\rm in}}=\sqrt[3]{7c_7/8c_4}T/\sigma\approx (0.50{\rm fm})T_{\rm^{TK}}. \label{RS}\ \ \ 
\end{eqnarray}

So far, we have considered the single cluster model.
In practice, the box wall should be taken as an approximation 
	to averaged effects of the other clusters in the environment.
The reflections by the wall are identified with the average effects of 
	the clusters outgoing and incoming through the boundary. 
The collisions may not be elastic, and the clusters may be 
	distorted, divided, and fused, and equilibrium would be incomplete.
The clusters are labeled by various quantum numbers as well as their macroscopic shapes and motions.
They are sifted through interactions among them and with the radiations, 
	and the suited survive the longer \cite{Darwin}.  
Now the parameter $L$ is the typical interval of clusters, 
	precisely defined by $L=\rho^{-1/3}$ with the number density $\rho$ of the clusters.

\section{consequences}

Now we assume the space expansion 
	of the Robertson-Walker-Friedmann type. 
Then, the scale factor of the space is known to be inversely proportional 
	to the temperature apart from small deviations of the coefficients 
	due to the changes of components responsible for the expansion. 
On the other hand, the baryon number is conserved
	in the period of our interest. 
Therefore, we roughly have for the baryon number density $\rho_{\rm B}$
\begin{eqnarray}&&
	\rho_{\rm B}/T^3\approx
	\rho_{\rm B0}/{T_0}^3\approx(10^{-11}\rm/fm^{3})/TK^{3} \approx10^{-10},\ \ \ \ \ \ 
\label{T3rho}
\end{eqnarray}
where 
	$T_0\approx2.7K$ is the present temperature of the cosmic microwave background,	
	and $\rho_{\rm B0}\approx 0.2/\rm m^3$ is 
	the baryon number density at present \cite{Weinberg}.

At very high temperature, we assume that everything was uniform 
	as the consequence of the cosmic inflation. 
As the space expanded, however, 
	the rare and discrete baryon numbers were separated piecewise
	yielding nonuniformity. 
The baryon number cores are expected to behave heterogeneously 
	as is supposed from the baryon spectra and interactions at low energy.
Then, this nonuniformity might give chances for the confining forces 
	to form quark-gluon clusters. 
The baryon numbers have no tendency to gather together
	as is seen from the fact that, at low energy, 
	no stable multi-baryon-number state directly due to QCD is observed
	other than those via nuclear forces due to meson exchanges.   
Since the baryon number was rare and macroscopically uniform,
	the most of the clusters are expected to have had $N_{\rm B}=1$.
Though clusters with $N_{\rm B}=0$ might also be formed, 
	for example through collisions,
 	they were unstable because of lack of protecting quantum number,
	and easily converted to radiations.
Hence, we identify the density $\rho$ of the clusters 
	with the baryon number density $\rho_{\rm B}$
	in the background of photon lepton plasma.

In FIG. 1 \cite{fig1}, %.\ \ref{fig1}, 
we show the ``phase diagram" in the $\rho_{\rm B}$-$T$ plane.
The upper boundary of the cluster phase (the dashed line in FIG. 1 \cite{fig1} %.\ \ref{fig1}) 
	is drawn at 
\begin{eqnarray}&&
	\rho_{\rm B}T^3\approx(4c_3/\pi^3c_6)\sigma^3,\ \ \mbox{ (at the boundary)}
\label{rhoT3}
\end{eqnarray}
where
	the extrapolations of $\langle E\rangle$ 
	in (\ref{EhiT}) and (\ref{EmidT}) coincide. 
In the view of (\ref{int}), this is not a sharp phase transition but a vague crossover effect.
Meanwhile, the cluster phase transits to the hadronic phase 
	at $T\approx T_{\rm had}$ where the chiral symmetry is broken, 
	quark-antiquark pairs are not produced, gluon production is also suppressed,
	and microscopic bound states due to gluon exchange become important.

\framebox{Fig.\ 1}\ \cite{fig1}

%\begin{figure}[h]  
% \centering
%  \includegraphics[width=8.3cm]{figure1.jpg}
% \caption{``Phase diagram" in the $\rho_{\rm B}$-$T$ plane.
%	The dashed line is the boudary of the ``phases", 
%	the solid line is the path of the early universe, and 
%	the shaded area is that of the heavy-ion collision experiments. }
% \label{fig1}
%\end{figure}

Now we draw the $\rho_{\rm B}$-$T$ curve (\ref{T3rho}) of the universe (the solid line) 
	on the diagram in FIG. 1 \cite{fig1}. %~\ref{fig1}. 
This indicates that the clusters began to be formed at
	$T\approx \rm7\times10TK\approx6GeV$ and $L\approx\rm6\times10fm$
	(where (\ref{T3rho}) and (\ref{rhoT3}) intersect)
	as the first composite entities in the universe.
Their typical size was $R\approx\rm3\times10fm$ by (\ref{R}),
	their average mass was $M\approx \rm8\times10^{11}GeV$ by (\ref{EmidT}),
	and they involved about $N\approx4\times\rm10^{10}$ quarks and gluons
	per a cluster on average by (\ref{NmidT}).
The size $R$, the mass $M$ and number $N$ decreased with the decreasing temperature $T$. 
They behaved like $R\approx 0.42{\rm fm}\,T_{\rm^{TK}}{}$ by (\ref{R}), 
	$M\approx 0.13{\rm GeV}\,T_{\rm^{TK}}{}^7$ by (\ref{EmidT}), 
	and $N\approx0.25\,T_{\rm^{TK}}{}^6$ by (\ref{NmidT}). 
On the other hand, with the decrease of the baryon density $\rho_{\rm B}$ by (\ref{T3rho}), 
	the typical cluster interval increased like $L\approx 5\times10^3{\rm fm}/T_{\rm^{TK}}{}$.
Since $R\ll L$, the system looks like a gas of clusters (except for its early stage). 
This exhibited the first macroscopic inhomogeneity in the universe 
	accompanied by the first information accumulation,
	as we see next.

Each cluster keeps its existence due to the facts that 
	its constituents are subject to color confinement of QCD, 
	it is protected by the baryon number, and it has appropriate energy in the given environment. 
The facts are taken as pieces of information to guarantee its existence.
The amount $I_{\rm qgc}$ of the information is obtained by $I_{\rm{qgc}}=S_0-S$, 
	where $S$ is the Shannon entropy 
	in (\ref{intermediate}) in the cluster phase, and
	$S_0=4c_4T^3L^3/3$ is that \underline{without} the information.
The latter ($S_0$) can be estimated 
	considering the model with the same matter contents
	but without gauge symmetry.
In practice, $S_0\gg S$, and we have 
	$I_{\rm qgc}\approx S_0\approx$ $1.7\times10^2\rm Gnat\approx2.5\times10^2Gbit$.
Owing to (\ref{T3rho}), this was almost constant while the universe is in the cluster phase. 
This is huge because the cluster is a macroscopic object.
It is interesting that  
	$i_{\rm qgc}\equiv I_{\rm qgc}/N_0=4c_4/3c_3\approx4.0\rm nat\approx5.7 bit$,
	where $N_0=c_3 T^3L^3$ is the average number of particles \underline{without} the information.

The information amount was mostly lost through breakdown of the chiral symmetry.
The clusters reduced to hadrons, and among them nucleons survived owing to the information they held. 
The nucleons inherited from the clusters the important information 
	that they are subject to color confinement and are protected by baryon number conservation, 
	although the amount a la Shannon was only a small part. 
Interestingly, the information is still inherited by all the sophisticated composite entities in the world around us today,
	since their existence is necessarily based on existence of the nucleons as their constituents \cite{Akama}.

Then, are such clusters formed at present after so long expansion and cooling of the universe? 
It becomes possible just because the inhomogeneity grew with local information accumulation. 
We can mention two possibilities: 
	one is due to gravitational localization and compression of matter, 
	and the other is due to human intelligence, 
	i.e.\ deep inside heavy stars and in scientific experiments by human beings. 
People claim that QGP is produced 
	in high energy collisions of heavy nuclei \cite{QGPrev}.
The ``QGP'' observed  
	was strongly interacting and looks rather like liquid, 
	differing from the naive expectation of QCD, as was a big discovery \cite{sQGP}.
This left, however, a big puzzle about its microscopic mechanism.    
The quark-gluon cluster may render a hint for the puzzle, as follows.

In the experimental setting,
	the baryon number is much denser than (\ref{T3rho}), 
	the temperature is lower,
	the relevant time is shorter, and 
	the equilibrium is incomplete. 
Roughly, the system is around $T\sim$ a few TK and $L\sim$O(fm),
	the shaded area in FIG. 1 \cite{fig1}, %.\ \ref{fig1}, 
	and rapidly transits from the plasma phase to cluster phase. 
Although our approximations turn less credible in this area,
	qualitative considerations are still useful.

In the early stage, it is in the plasma phase near the boundary, 
	where the confining interactions become significant compared with the kinetic energies. 
Owing to the confining force, 
	the quarks and gluons behave in loose groups of the size O($T/\sigma$).
The groups are nothing but the precursors of the quark-gluon clusters.
By the (classical) dimensional analysis, their mean free path is of O($T/\sigma$),
	and the system looks like liquid.
If we dare to use (\ref{NmidT})--(\ref{RS}), $R\sim$ O(fm), $M\sim$ O(GeV's), and $N\sim$ O($10^2$)
	while $L$ is of O(fm).
Therefore. the clusters are densely distributed, the mean free path is also of O(fm),
	and hence interact strongly through their direct contacts.
Thus, throughout the process, the system is strongly interacting, and looks like liquid. 
This is in conformity with the experimental observations \cite{QGPrev}, \cite{sQGP},
	and would solve the puzzle stated above. 
The strong interactions come from confinement-driven clustering of quarks and gluons. 
We expect similar behaviors in the quark-gluon clusters 
	around the QGP in the dense stars, if any. 
Theoretical and experimental investigations are desired.

\section{Conclusions}

We have incorporated effects of color confining potential $U(r)=\sigma r$ 
	into statistical formulation by counting its energy levels.
This indicates emergence of the macroscopic quark-gluon clusters.
Taking into account the space expansion and the baryon-number conservation \cite{Weinberg}, 
	we conclude that such clusters 
	existed in the early universe in the period when $T_{\rm had}\lesssim T\lesssim70 \rm TK$.
Their typical size $R$, their average mass $M$, the average number $N$ of quarks and gluons in a cluster, 
	and the typical interval $L$ of the clusters were 
	$R\approx 0.42{\rm fm}\,T_{\rm^{TK}}$, 
	$M\approx 0.13{\rm GeV}\,T_{\rm^{TK}}{}^7$, 
	$N\approx0.25\,T_{\rm^{TK}}{}^6$, and 
	$L\approx 5000{\rm fm}/T_{\rm^{TK}}$
	with $T_{\rm^{TK}}=T\rm/TK$. 
The universe in the period looked like a gas of the clusters 
	and exhibited the first macroscopic inhomogenuity in the universe. 

The quark-gluon clusters held a large amount of information, 250Gbit on average, for their own existence. 
This is constant while the universe is in the cluster phase. 
Though this was the first information realized in nature,
	the amount was mostly lost through braekdown of the chiral symmetry.
The clusters finally reduced to nucleons, 
	which are subject to confinement and are protected by baryon number conservation.
The nucleons inherited this information from the clusters, 
	though the amount a la Shannon was only a tiny part.
The information is still inherited by all the sophisticated composite entities 
	in the world around us today \cite{Akama}.

On the other hand, in the heavy ion collision experiments,
	the temperature and the baryon number density of the ``QGP"
	are near the boundary of the cluster and the QGP phases. 
Then, the clusters (or their precursors)  interact strongly and the system qualitatively looks like liquid.
This is in conformity with the experimental observations of QGP \cite{QGPrev}, \cite{sQGP}.
This would give a microscopic answer to the question 
	why the QGP in the experiments looks like liquid unlike the naive QCD expectation.
These facts may support existence of such clusters there.

The approximations used here leave much room for improvements. 
For example, the mean field potential, 
	the temperature dependence of the baryon density in the universe,
	and the treatments around the phase boundary 
	would be replaced by more concrete and sophisticated methods.
We believe, however, the gross features would not much be altered
	and would have a certain extent of significance.

\acknowledgements
The author would like to thank 
Dr.~H.~Fujii,
Dr.~T.~Hattori, Dr.~Y.~Kawamura, Dr.~H.~Mukaida, 
Dr.~S.~Ono, Dr.~S.~Suzuki, Dr.~S.~Wada,
and Dr.~S.~Yazaki for discussions.


\begin{thebibliography}{99}
\bibitem{composite}
We follow the presently established standard view that  
	the fundamental matters are the quarks, the leptons, the gauge bosons,
	and the Higgs bosons \cite{standard},
	although we cannot exclude their compositeness \cite{subquark}.
\end{thebibliography}
\end{document}